\documentclass[twocolumn,showpacs,prb]{revtex4}% Physical Review B
\usepackage{graphicx}% Include figure files
\usepackage{dcolumn}% Align table columns on decimal point
\usepackage{bm}% bold math
\usepackage{amsmath}
\usepackage{amssymb}
\usepackage{relsize}

\newcommand{\CQCTsyd}{Australian Research Council Centre of Excellence for Quantum Computer Technology, \\
School of Electrical Engineering \& Telecommunications, University of New South Wales, Sydney NSW 2052, Australia.}

\newcommand{\TKKcomp}{Department of Applied Physics/COMP, Aalto University, P.O.~Box 15100, FI-00076 AALTO, Finland.}
\newcommand{\TKKltt}{Low Temperature Laboratory, Aalto University, 
  P.O.~Box 13500, FI-00076 AALTO, Finland.}
\newcommand{\smallsum}{\ensuremath{\mathsmaller{\sum}}}

\begin{document}

%\nofiles
\title{Non-Abelian geometric phases in ground state Josephson devices}

%\preprint{APS/123-QED}

\author{J.-M.~Pirkkalainen}
\affiliation{\TKKcomp} \affiliation{\CQCTsyd}

\author{P.~Solinas}
\affiliation{\TKKcomp}

\author{J.~P.~Pekola}
\affiliation{\TKKltt}

\author{M.~M\"{o}tt\"{o}nen}
\affiliation{\TKKcomp} \affiliation{\CQCTsyd} \affiliation{\TKKltt}

\date{\today}

\begin{abstract}
We present a superconducting circuit in which non-Abelian geometric transformations can be realized using an adiabatic parameter cycle. In contrast to previous proposals, we employ quantum evolution in the ground state. We propose an experiment in which the transition from non-Abelian to Abelian cycles can be observed by measuring the pumped charge as a function of the period of the cycle. Alternatively, the non-Abelian phase can be detected using a single-electron transistor working as a charge sensor.
\end{abstract}

%\pacs{03.65.Vf, 85.25.Cp}% PACS, the Physics and Astronomy

\maketitle

% Introduction
%{\bf Introduction - 1 page}

{\it Introduction} --- Accurate control and measurement of few-level quantum systems has recently attracted great experimental and theoretical interest with possible applications in quantum information processing (QIP). Geometric phases~\cite{Anandan1997} arising from adiabatic and cyclic quantum evolution can provide robustness against, e.g., timing errors. Recently, it was shown that such evolution in a non-degenerate ground state is immune to decoherence from a low-temperature environment~\cite{Pekola2009} suggesting that it may provide an important tool for controlling quantum systems.

In the non-degenerate case, the accumulated geometric phase, the Berry phase~\cite{Berry1984}, is a shift of the complex phase of the eigenstate, and hence cannot be used as such for QIP. Non-Abelian phases~\cite{Simon1983,WilczekZee1984} correspond to unitary matrices operating in a degenerate subspace of the system Hamiltonian, thus providing means for universal QIP~\cite{ZanardiRasetti1999}. Although geometric phases capable of entangling two quantum bits, qubits, have been observed in liquid-state nuclear magnetic resonance experiments~\cite{Jones2000}, this kind of geometric quantum computing (GQC) is yet to be demonstrated. In fact, the geometric phases using nuclear magnetic resonance~\cite{Jones2000}, and in more recent experiments~\cite{Morton2005} demonstrating non-adiabatic Aharonov-Anandan phases~\cite{Aharonov1987} in fullerene spin qubits, accumulate in a rotating frame, and hence there is no true degeneracy in the system.

%%%%%%%%%%%%%%%%%%%%%%%%%%%%%%%%%
% New Paolo Introduction begins %
%%%%%%%%%%%%%%%%%%%%%%%%%%%%%%%%%

The initial proposals for the experimental realization of GQC~\cite{Unanyan1999,DuanCiracZoller2001} rely on a fully degenerate subspace to build the logical operators and it has been extended to many quantum systems~\cite{Fuentes2002,Recati2002,Solinas2003} including Josephson junction devices~\cite{Faoro2003,Choi2003}. In similar systems, a way to observe the non-Abelian evolution by measuring the charge pumped through the device has been recently proposed~\cite{Brosco2008}.

\begin{figure}[th!] \center
%\texttt{.pdf}

\includegraphics[width=0.99\linewidth]{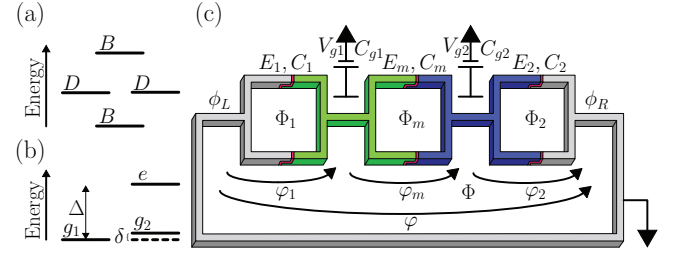}
\caption{ \label{fig1} (color online) (a) Eigenenergy diagram of the so-called tripod Hamiltonian consisting of two bright states, $B$, and two degenerate dark states, $D$. (b) Eigenenergy diagram of the circuit Hamiltonian along the considered cycle. The energy difference between the ground, $g$, and excited, $e$, state is denoted by $\Delta$ and the ground state degeneracy splitting by $\delta$. In the ideal case, $\delta=0$. (c) Schematic diagram of the Non-Abelian superconducting pump. Green and blue denote the superconducting islands and red the Josephson junctions.}
\end{figure}

However, all the schemes assume typically a so-called tripod Hamiltonian which has degeneracy only in its excited states [see Fig.~\ref{fig1}(a)] rendering the system prone to decoherence even in the low-temperature limit. This is potentially a serious limitation in the condensed matter systems in which the coupling between system and environment is strong and unavoidable.

In this paper, we present an experimentally realizable Josephson device and show that it can be used to observe adiabatic non-Abelian geometric phases. In contrast to the above pioneering works, we employ a conceptually different Hamiltonian allowing us to work on the ground state manifold of the system. This proposal provides a clear extension to the theoretical proposals~\cite{Pekola1999,Aunola2003,Mottonen2006} and experiments~\cite{Leek2007,Mottonen2008} on the Berry phase in superconducting circuits.

{\it Non-Abelian adiabatic evolution} --- We denote the parameters
of the system Hamiltonian in a general cyclic loop by a vector
$\vec{\lambda}$. The instantaneous eigenstates of the Hamiltonian
$H[ \vec{\lambda}(t) ]$ along this loop for all $t \in [0,T]$ are
denoted by $\{ | \psi_{\alpha}(t) \rangle \}$, where $T$ is the
period of the cycle. Generally, any temporal evolution of the system
state can be represented using the time evolution operator, $U(T)$,
such that $| \Psi(T) \rangle = U(T) | \Psi(0) \rangle$, where $ |
\Psi(t) \rangle $  is the state of the system at time $t$. The
charge $Q$ transferred through a superconducting system in one
parameter cycle can be obtained by integration of the current
operator $\hat{I}= -\frac{2e}{\hbar} \partial_{\varphi} H(t)$ as
%\begin{equation}
$Q = \int_0^T \mathrm{d}t \, \langle {\Psi}(t) |
\hat{I} | {\Psi}(t) \rangle$,
%\end{equation}
where and $\varphi$ is the superconducting phase difference across
the system~\cite{Brosco2008,Mottonen2008}. Using the Schr\"odinger
equation and the definition of the time evolution operator, this can
be written in the form

\begin{equation}
Q = - 2ie \langle \Psi(0) | U^{\dagger}(T) \left[ \partial_{\varphi}
U(T) \right] | \Psi(0) \rangle. \label{eq:PumpedSimple}
\end{equation}

However, if the Hamiltonian parameters are changed adiabatically
along the cycle, the evolution can be restricted to the initial
eigenspace. In an $n$-fold degenerate eigenspace, the state of the
system after a parameter cycle is
%\begin{equation}
$| \Psi(T) \rangle = U_{\mathrm{ad}}(T) | \Psi(0) \rangle +
\mathcal{O}(1/T)$,
%\label{eq:TimeEvolution}
%\end{equation}
where $ |{\Psi}(t) \rangle = \sum_{i=1}^n c_i(t) | \psi_i(t) \rangle
$~\cite{WilczekZee1984}. If the instantaneous eigenvectors are
defined globally and continuously, the operator $U_{\mathrm{ad}}(t)$
is represented in this basis as

\begin{equation}
U_{\mathrm{ad}}(t) = e^{-(i/\hbar) \int_0^t \mathrm{d}t' E(t')} \mathcal{T} e^{- \int_0^t \mathrm{d}t' \Gamma (t')},
\label{eq:Holonomy}
\end{equation}
where $E(t)$ is the energy of the degenerate eigenspace,
$\mathcal{T}$ is the time ordering operator, and the connection
$\Gamma (t)$ is given by $\left[ \Gamma (t)  \right]_{\alpha \beta}
= \langle \psi_{\alpha} (t) | \dot{\psi}_{\beta} (t) \rangle$. The
first exponential function in Eq.~(\ref{eq:Holonomy}) yields the
accumulated dynamic phase shift, $U_{\mathrm{dyn}}(t)$, and the second one
provides the geometric transformation, $U_{\mathrm{geo}}(t)$, which
is non-Abelian in general.

In the adiabatic limit, $U(T)$ can be replaced with
$U_{\mathrm{ad}}(T)$, and substituting Eq.~(\ref{eq:Holonomy}) into
Eq.~(\ref{eq:PumpedSimple}) yields the relation between the
different transformations and the transferred charge

\begin{align}
Q = - 2ie \Big( & \langle \Psi(0) | U_{\mathrm{geo}}^{\dagger}(T)
\left[ \partial_{\varphi} U_{\mathrm{geo}}(T) \right] | \Psi(0)
\rangle \nonumber \\ + & \langle \Psi(0)
| U_{\mathrm{dyn}}^{\dagger}(T) \left[ \partial_{\varphi}
U_{\mathrm{dyn}}(T) \right] | \Psi(0) \rangle \Big) \label{eq:PumpedConnection},
\end{align}
where the first term is the geometric pumped charge and the second the dynamic charge due to the usual supercurrent. In the case of a nondegenerate eigenspace, $n=1$, this reduces to the well known relation $Q = 2e \partial_{\varphi} (\Theta_B -\Theta_d)$, where the accumulated Berry phase, $\Theta_B$, is related to $U_{\mathrm{geo}}$ by $U_{\mathrm{geo}} = e^{i \Theta_B}$ and the dynamic phase, $\Theta_d$, to $U_{\mathrm{dyn}}$ by $U_{\mathrm{dyn}} = e^{-i \Theta_d}$~\cite{Mottonen2006}. See Brosco \emph{et al.}~\cite{Brosco2008} for an alternative way to obtain the pumped charge. Although the Berry phase induces just a phase shift to the state vector, it does not commute in general with the current operator $\hat{I}$ which originates from a higher-dimensional system.

% * Starting from the Eq. (1) in Brosco paper, we obtain the pumped current in terms of the geometric transformation:

% * Cyclic transformation yields the equation:

% * The equation above still haves both the supercurrent and geometric contribution

% * In our cycle, the supercurrent is zero in the ideal case

% Description of the device
%{\bf Device description - 0.5 pages}

{\it Model circuit} --- The Cooper pair pump shown in Fig.~\ref{fig1}(c) is considered here as the physical realization for observing non-Abelian geometric phases. It consists of three SQUIDs in series with two superconducting islands between them. The SQUIDs are operated as tunable Josephson junctions which can be closed (Josephson energy $E_i$ is zero) and opened ($E_i \neq 0$) by controlling the magnetic flux through them. The phase difference of the order parameter across the whole device, $\varphi = \phi_L - \phi_R$, is kept constant by the magnetic flux $\Phi$ through the outermost loop. The Hamiltonian has five external parameters which are controlled during a pumping cycle, i.e., three magnetic fluxes and two gate voltages.

%It is a combination of the so-called sluice pump and the typical Cooper pair pump employed to pump current adiabatically in Refs.~\cite{Mottonen2008} and~\cite{Geerligs1991}, respectively.

%%%%%%%%%%%%%%%%%%%%%%%%%%%%%%%%%%
%%% These equation can be cut %%%%
%%%%%%%%%%%%%%%%%%%%%%%%%%%%%%%%%%
The charging energy part of the Hamiltonian, $H_{\mathrm{ch}}$, is given by

\begin{align}
H_{\mathrm{ch}} = & E_{C_1} (\hat{n}_1 - n_{g1})^2 + E_{C_2} (\hat{n}_2 - n_{g2})^2 \nonumber \\ + & E_m (\hat{n}_1 - n_{g1}) (\hat{n}_2 - n_{g2}),
\end{align}
where $\hat{n}_i$ is the operator for the excess number of Cooper pairs on the $i$th island and $n_{gi}$ is the corresponding gate charge given by $n_{gi} = C_{gi} V_{gi} /(2e)$. The charging energies are $E_{C_1} = 2 e^2 C_{\smallsum_2}/C^2$, $E_{C_2} = 2 e^2 C_{\smallsum_1}/C^2$, and $E_m = 4 e^2 C_m/C^2$. Here, $C_{\smallsum_i}$ is the total capacitance of the $i$th island and $C^2 = C_{\smallsum_1} C_{\smallsum_2} - C_m^2$.

The Josephson part of the Hamiltonian, $H_J$, reads

\begin{align}
H_J = \sum_{n_1,n_2=-\infty}^{\infty} \Big( & J_{\mathrm{eff},1} |n_1+1,n_2 \rangle \langle n_1,n_2| \nonumber \\
+ & J_{\mathrm{eff},m} |n_1+1,n_2-1 \rangle \langle n_1,n_2| \nonumber \\
+ & J_{\mathrm{eff},2} |n_1,n_2+1 \rangle \langle n_1,n_2| + \mathrm{h.c.}  \Big),
\end{align}
where $|n_1,n_2 \rangle$ denotes the state with $n_i$ excess Cooper pairs on the $i$th island, $J_{\mathrm{eff},1} = -E_1(\Phi_1) e^{i \varphi(\Phi)/2} /2$, $J_{\mathrm{eff},2} = -E_2(\Phi_2) e^{-i \varphi(\Phi)/2} /2$, and $J_{\mathrm{eff},m} = -E_m(\Phi_m)/2$. Here, $E_1$, $E_2$, and $E_m$ are the tunable Josephson energies. The full Hamiltonian is given by $H = H_{\mathrm{ch}}(V_{g1},V_{g2}) + H_J(\Phi_1,\Phi_2,\Phi_m,\Phi)$.

\begin{figure}[th!] \center
%\texttt{.pdf}

\includegraphics[width=0.99\linewidth]{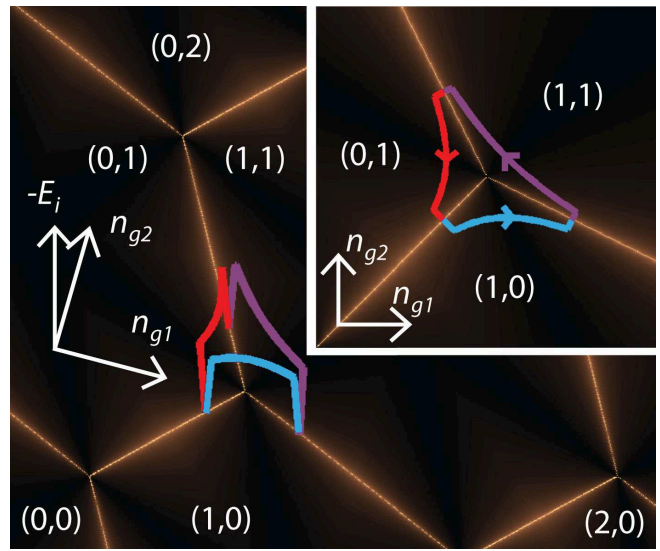}
\caption{ \label{fig2} (color online) Parameter cycle inducing non-Abelian transformations with the stability diagram as the background. The charge state $|i,j\rangle$ which minimizes the charging energy in each hexagonal area is denoted by ($i,j$) The $z$ axis represents $E_i$ with blue in the front for $E_1$, purple on the right for $E_m$, and red on the left for $E_2$. The SQUID energies can be closed down to $E_i^{\mathrm{min}}$, which is zero in the ideal cycle. The inset shows the parameter cycle projected into the gate voltage plane.}
\end{figure}

{\it Non-Abelian cycle} --- If all the SQUIDs are closed, the conventional stability diagram with a hexagonal structure is recovered~\cite{LeoneLevy2008}, see Fig.~\ref{fig2}. In the vicinity of the triple degeneracy point of states $|1,0\rangle$, $|0,1\rangle$, and $|1,1\rangle$, the adiabatic evolution is approximately restricted to these three states. The parameter cycle is composed of three symmetric paths in each of which a SQUID is opened, the gate voltages are shifted along a ground state degeneracy, and finally the SQUID is closed.

Along each path, the effective three-level Hamiltonian has a 2\,$\times$\,2 block and can be written as 
$H_{\mathrm{eff}}=\vec{\sigma}_{i,j} \cdot \vec{B}(t) + \epsilon_k(t) |k\rangle \langle k|$, where $\vec{\sigma}_{i,j}= \{\sigma^x_{i,j},\sigma^y_{i,j},\sigma^z_{i,j} \} $ is a vector composed of the Pauli matrices for the states $i,j$ (for example,  $\sigma^x_{i,j}= |i\rangle \langle j|+ |j\rangle \langle i|$), $\vec{B}(t)$ is an effective magnetic field, $\epsilon_k$ is the eigenvalue of the third charge state, and $\{|i\rangle,|j\rangle,|k\rangle\} = \{|1,0\rangle,|0,1\rangle,|1,1\rangle\}$.

The condition of the ground state double degeneracy is satisfied by tuning the smaller eigenvalue of the 2\,$\times$\,2 block of $H_{\mathrm{eff}}$ to be equal to $\epsilon_k(t)$ along the evolution. In the three-level approximation this implies that the degenerate gate voltage paths are hyperbolas in the gate voltage plane with one SQUID kept open. Along the opening and closing of the SQUIDs, we choose to change voltages linearly with the SQUID energies. In this way, a nontrivial loop encircling the triple degeneracy point can be traversed along a path with a doubly degenerate ground state.

Using the eigenstates along the three paths, we can construct a continuous global basis (defined in the whole parameter space) and calculate the connection $\left[ \Gamma (t)  \right]_{\alpha \beta}$. If the SQUIDs can be perfectly closed, the supercurrent due to the dynamic phase in Eq.~(\ref{eq:PumpedConnection}) vanishes since the energies of the eigenstates do not depend on $\varphi$. In this case, the transferred charge has only a geometric contribution which can be calculated analytically from the $\varphi$ dependence of the $U_{\mathrm{geo}}(T)$ operator. For a cycle starting from the degeneracy line between the states $|1,0\rangle$ and $|0,1\rangle$, this yields for the geometric transformation

\begin{equation}
U_{\mathrm{geo}}(T) = \left[ \begin{array}{cc}
0 & e^{i \varphi} \\
1 & 0
\end{array} \right], \label{eq:NAholonomy}
\end{equation}
represented in the basis $\{ |1,0\rangle,|0,1\rangle \}$. This result was confirmed by solving numerically the Schr\"odinger equation using 25 charge states indicating that our analysis does not rely on the three-state approximation. The obtained transformation is topological in the sense that it does not depend on the exact values to which the SQUIDs are opened as long as the evolution is kept degenerate along the cycle. From Eq.~(\ref{eq:PumpedConnection}), we obtain for the geometrically pumped charge

\begin{equation}
Q = 2e \left[ c_1^*(0), c_2^*(0)  \right] \left[ \begin{array}{cc}
0 & 0 \\
0 & 1
\end{array} \right] \left[ \begin{array}{cc}
c_1(0)  \\
c_2(0)
\end{array} \right] = 2e |{c_2(0)}|^2. \label{eq:pumped}
\end{equation}
Thus, the pumped geometric charge is independent of the phase across the device and depends only on the initial state.

% If the other triple degeneracy is traversed:

%\begin{equation}
%U_{\varphi}(T) = \left[ \begin{array}{cc}
%0 & 1 \\
%e^{i \varphi} & 0
%\end{array} \right]. \label{eq:NAholonomy_2}
%\end{equation}

%  The straight lines of the parameter cycle denote the parts of the cycle in which the SQUIDs are being opened or closed.

% * Need to say that the $E_c$'s were fixed to this and that for the shape of the parameter cycle to make sense (depends on the actual values).

% Measuring the abelianity
%{\bf Observing non-Abelianity - 1,5 pages}

{\it Observation scheme for non-Abelian phases} --- Here we discuss two methods to observe the non-Abelian transformations. Firstly, a single-electron transistor (SET) can be coupled asymmetrically to the superconducting islands and used as a charge sensor. The additional capacitance due to the SET changes slightly the charging energies but does not affect the operation principle of the circuit. Initializing the system and performing the parameter cycle adiabatically swaps the charge states of the islands regardless of the phase across the device, which can be detected with the charge sensor. Observation of this charge transfer proves the non-Abelian character of the evolution since in the Abelian case, initial populations are conserved in cyclic adiabatic evolutions.

Another way to observe the non-Abelian features is to measure the pumped charge through the system using a detector junction~\cite{Mottonen2008}. Since in the experiments the SQUIDs cannot be perfectly closed~\cite{Faoro2003,Kemppinen2008}, we consider here a case in which the Josephson energies can be tuned down to $E_i^{\mathrm{min}}$ of their maximum value $E_i^{\mathrm{max}}$.% For the charging energy $E_C$ we use the value $E_C$ = 0.2 meV routinely realized in experiments~\cite{Mottonen2008}.

In the case of non-ideal SQUIDs, two additional effects have to be considered. Firstly, the supercurrent contribution usually dominates over the geometric contribution. However, it has been shown~\cite{Mottonen2008} that the supercurrent contribution can be efficiently measured by traversing the parameter cycle first forwards and then backwards. In the perfect adiabatic limit, the geometric component of the current cancels itself and the measured total current is twice the supercurrent.

%Experimentally, the supercurrent component
%can be distinguished by traversing the parameter cycle first
%forwards and then backwards. In the perfectly adiabatic limit, the
%geometric component of the current cancels itself and the measured
%total current is twice the supercurrent. Subtracting the
%supercurrent from the total current obtained from single cycle
%manipulation yields the pumped charge.

Secondly, the two lowest-energy eigenstates are not perfectly degenerate with the energy gap $\delta \sim E_i^{\mathrm{min}}$, see Fig.~\ref{fig1}(b). To obtain non-Abelian evolution, the loop has to be traversed fast enough such that the two lowest eigenstates are effectively degenerate, that is $T \ll \hbar/\delta$ where $T$ is the cycle period. On the other hand, the evolution should be slow enough to avoid transitions to the higher states implying $T \gg \hbar/\Delta$, where $\Delta$ is the energy gap to the excited state. To obtain the Abelian limit, the energy gap $\delta$ can be increased by larger Josephson energies and the cycle can be traversed slower such that no transitions occur.

The system can be initialized to the state $|1,0 \rangle$ by the following procedure. First, all the SQUIDs are closed to $E_i^{\mathrm{min}}$ and gate voltages tuned to have $|1,0 \rangle$ as a nondegenerate ground state. After the system has relaxed to the ground state, the gate voltages are suddenly shifted, $T_{\mathrm{shift}} \ll \hbar/\delta$, to the degeneracy line between the states $|1,0 \rangle$ and $|0,1 \rangle$. The sudden shift keeps the system in the state $|1,0 \rangle$ and the non-Abelian cycle can be traversed starting from a well-known initial state. The system can be initialized to the state $|0,1 \rangle$ with a similar procedure.

% A similar procedure can be used to initialize the system to the state $|0,1 \rangle$.

To describe the adiabaticity of the evolution, we introduce the adiabaticity parameter $\alpha$ defined as the population of the initial state after a back-and-forth cycle. In the perfectly non-Abelian regime, the geometric transformations induced by the forward and backward cycles exactly cancel each other. Thus, the total transformation is proportional to the identity implying that $\alpha=1$. For the perfectly Abelian limit, no transitions occur between the eigenstates and again $\alpha=1$ if the initial state is an eigenstate. Between these two regimes, no easy theoretical prediction can be made since the states are only partially mixed during the evolution.% In principle, adiabatic regimes could be found with faster cycles in which three or more lowest eigenstates are perfectly mixed and adiabaticity would again be unity but this was not the case with the cycle considered here. 

\begin{figure}[th!] \center

\includegraphics[width=0.99\linewidth]{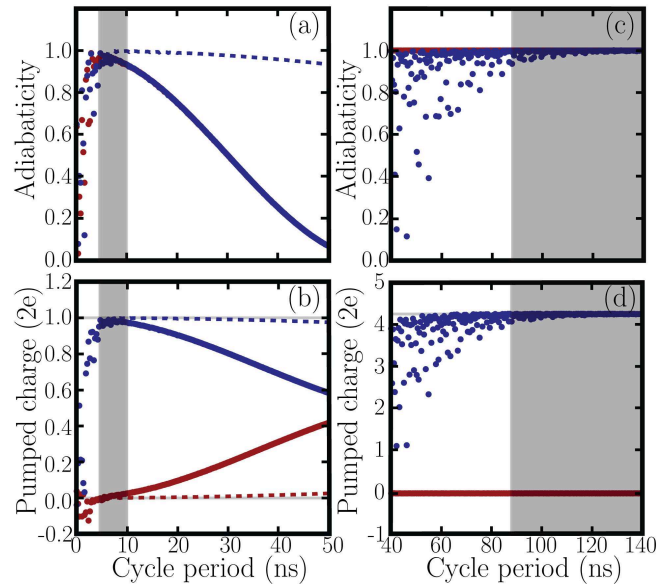}
\caption{ \label{fig3} (color online) (a) Adiabaticity as a function of the cycle period in a cycle with $E_i^{\mathrm{max}} = -0.4 E_C$ and $E_i^{\mathrm{min}} = E_i^{\mathrm{max}}/k$. Red (blue) denotes a cycle starting from the initial state $|1,0 \rangle$ ($|0,1 \rangle$) for $k=1000$ (dots) and $k=5000$ (dashed line). (b) The pumped charges corresponding to (a). (c) Adiabaticity of the ground (red) and excited state (blue) in a cycle with all the SQUID energies fixed to $E_i^{\mathrm{max}}=E_i^{\mathrm{min}}=-0.4 E_C$. (d) The pumped charges corresponding to (c). In (b) and (d) the gray lines denote the geometric pumped charges in the perfectly adiabatic limits. The shaded areas indicate the cycle periods with which the evolution is adiabatic. The charging energy $E_{C_i}$ used is experimentally realizable 0.2 meV~\cite{Mottonen2008} and the phase across the device is fixed to zero, $\varphi=0$.}
\end{figure}

In all numerical simulations, we fix the phase across the device $\varphi$ to zero and $E_{C_i}$ = 0.2 meV. Figure~\ref{fig3}(a) suggests that for non-Abelian cycle with period 5 ns $\leq T \leq$ 10 ns the evolution is adiabatic and $\alpha$ is close to unity even if the SQUIDs cannot be perfectly closed with $k = E_i^{\mathrm{max}}/E_i^{\mathrm{min}}=1000$. In this regime, the pumped charge shown in Fig.~\ref{fig3}(b) reaches the value $2e$ or 0 depending on the initial state as predicted by Eq.~(\ref{eq:pumped}). With $k=5000$ the adiabatic evolution window is broad and observed as a pumped charge plateu. To obtain a measurable current with a reasonable averaging time ($>$ 1 pA)~\cite{Mottonen2008}, the pumping cycle needs to be repeated fast enough. If simply a sequence of repeated pumping cycles is performed, the measured current reflects the average pumped charge $e$ regardless of the initial state due to the swapping between the states $|0,1 \rangle$ and $|1,0 \rangle$. On the contrary, the system can be initialized to the same state before every cycle. In this case, the pumped charge per cycle is $2e $ or $0$ depending on the initial state. Measuring such dependence on the initialization indicates that the charge states are swapped after each cycle providing a fingerprint of the non-Abelian evolution.

%Numerical simulations suggest that for cycle period 5 ns $\leq T \leq$ 10 ns, the evolution along the cycle is adiabatic and $\alpha$ is close to unity even if the SQUIDs cannot be perfectly closed with $k = E_i^{\mathrm{max}}/E_i^{\mathrm{min}}=1000$, see Fig.~\ref{fig3}(a). 

The evolution can be made Abelian by increasing the cycle period and keeping all the SQUIDs constantly open with $E_i^{\mathrm{max}}=E_i^{\mathrm{min}}=-0.4 E_C$. Figure~\ref{fig3}(c) indicates that the evolution is adiabatic with cycle periods longer than $\sim$85 ns and additional results (not shown here) confirm that the evolution is Abelian. Numerical simulations for the pumped charge, shown in Fig.~\ref{fig3}(d), yield $8.5 e $ or $0$ depending on the initial state which are the two lowest eigenstates. The pumped charge in the Abelian limit depends on $\varphi$ which in the simulation is fixed to zero. Here, the two different procedures (with and without initialization between the pumping cycles) lead to the same average pumped charge pointing out that no swapping between the lowest eigenstates takes place.

%By repeating the initialization procedure and performing the parameter cycle, measurable currents ($>$ 1 pA) can be produced through the system from which the charge pumped by an individual parameter cycle can be deduced. In the non-Abelian pumping limit, measuring the pumped charge of two consecutive cycles and comparing it with the pumped charge of a single cycle reveals that the charge state is changed. Since the charge states are swapped after the first cycle, the total pumped charge of two consecutive cycles is the sum of pumped charges of initial states $|1,0 \rangle$ and $|0,1 \rangle$ regardless of the initial state. On the other hand, pumped charge by a single cycle depends strongly on the initial state according to Eq.~(\ref{eq:pumped}). Thus the non-Abelian character of the parameter cycle can be observed without a charge detector by measuring the pumped charge through the device.

%* a plot of the phase dependency of the pumped charge in the perfect Abelian case (slow limit, would give more easily distinguishable regions between Abelian and non-Abelian case)? (Fig. 5) This would also explain the oscillations in the SET technique briefly mentioned.

% Conclusions
%{\bf Conclusions - less than 0.25 pages}

In conclusion, we have presented a rather simple superconducting circuit with which non-Abelian geometric transformations can be realized in the ground state of the system. A parameter cycle is introduced for which the corresponding geometric transformation is determined analytically. Two observation schemes are presented for the non-Abelian features taking into account the most important experimental restrictions.

The authors thank  V.\ Pietil\"a for insightful discussions. This work is supported by Academy of Finland, Emil Aaltonen Foundation, and Finnish Cultural Foundation. This work was partially funded by the Australian Research Council, the Australian Government, the U.S. National Security Agency, the U.S. Army Research Office (under Contract No. W911NF-04-1-0290), and European Community's
Seventh Framework Programme under Grant Agreement No. 238345 (GEOMDISS).

\bibliography{manu}

\end{document}